\documentclass[12pt,preprint]{aastex}
\usepackage{amsmath,amssymb,amsfonts,natbib}
\def\cm{{\rm\,cm}} 
\def\m{{\rm\,m}}

\def\g{{\rm\,g}}

\def\mumoon{\mu_\mathrm{moon}}
\def\muring{\mu_\mathrm{ring}}
\def\muend{\mu_\mathrm{end}}
\def\tdiff{t_\mathrm{delay}}
\def\Plib{P_\mathrm{lib}}

\begin{document}

\title{Care and feeding of frogs}

\author{Margaret Pan\altaffilmark{1}, Eugene Chiang\altaffilmark{1,2}}

\altaffiltext{1}{Department of Astronomy, University of California,
    Berkeley, CA 94720}
\altaffiltext{2}{Department of Earth and Planetary Science, University of California,
    Berkeley, CA 94720}
    
\email{mpan@astro.berkeley.edu}

\begin{abstract}
  ``Propellers'' are features in Saturn's A ring associated with
  moonlets that open partial gaps. They exhibit non-Keplerian motion
  \citep{tiscareno10}; the longitude residuals of the best-observed
  propeller, ``Bl\'{e}riot,'' appear consistent with a sinusoid of
  period $\sim$4~years. \cite{pan10} proposed that propeller moonlets
  librate in ``frog resonances'' with co-orbiting ring material. By
  analogy with the restricted three-body problem, they treated the
  co-orbital material as stationary in the rotating frame and
  neglected non-co-orbital material. Here we use simple numerical
  experiments to extend the frog model, including feedback due to the
  gap's motion, and drag associated with the Lindblad disk torques
  that cause Type I migration.  Because the moonlet creates the gap,
  we expect the gap centroid to track the moonlet, but only after a
  time delay $\tdiff$, the time for a ring particle to travel from
  conjunction with the moonlet to the end of the gap. We find that
  frog librations can persist only if $\tdiff$ exceeds the frog
  libration period $\Plib$, and if damping from Lindblad torques
  balances driving from co-orbital torques.  If $\tdiff \ll \Plib$,
  then the libration amplitude damps to zero.  In the case of
  Bl\'{e}riot, the frog resonance model can reproduce the observed
  libration period $\Plib \simeq 4$ yr. However, our simple feedback
  prescription suggests that Bl\'{e}riot's $\tdiff \sim 0.01 \Plib$,
  which is inconsistent with the observed libration amplitude of
  260~km.  We urge more accurate treatments of feedback to test the
  assumptions of our toy models.
\end{abstract}

\section{Introduction}
\label{sec:introduction}

``Propellers'' observed by the {\em Cassini} spacecraft in Saturn's A
ring appear as S-like features superimposed on azimuthally long and radially
narrow gaps
\citep{tiscarenoetal06,srem07,tiscarenoetal08,tiscareno10}. Each propeller
is believed to trace a moonlet several hundred meters in size
which gravitationally repels nearby ring particles, creating an underdense
gap in the moonlet's immediate vicinity, as well as overdensities at
radii just outside the gap. Because of Keplerian shear, these density
perturbations propagate toward greater longitudes inside the moonlet's
orbit and smaller longitudes outside, producing two lobes separated by
a few Hill spheres of the moonlet \citep{sei05,lewis09}. The gap's
azimuthal length is set by the moonlet mass and the time for ring
particles to diffuse back into the gap via particle-particle
interactions \citep{spahn00,srem02,sei05,lewis09}.

Propellers in the outer A ring exhibit non-Keplerian motion
\citep{tiscareno10}. In particular, the longitude residuals of the
propeller ``Bl\'{e}riot,'' observed 89 times over 4.2~years, show
variations consistent with a sine wave of half-amplitude 260~km and
period 3.68~years. The longitude residuals imply semimajor axis
variations of order 100~m. These data constrain the underlying
mechanism. The variations' smooth
sinusoidal character suggests that this mechanism is not
stochastic on timescales shorter than a few years. Secular
interactions with the Saturnian moons, rings, or equatorial bulge have
timescales typically much longer than a few years; in any case, secular
interactions cannot induce semimajor axis changes. Finally, no other
Saturnian moon appears to occupy a mean motion resonance with
Bl\'{e}riot \citep{tiscareno10}.

Pan \& Chiang (2010, hereafter PC) proposed that Bl\'eriot's
non-Keplerian motion is caused by gravitational interactions between
Bl\'{e}riot's moonlet and co-orbital ring material outside the
moonlet's gap.  The interaction is long-range: the co-orbital mass is
located thousands of Hill sphere radii away from the moonlet. The
propeller moonlet and the much more massive co-orbital material
participate in a 1:1 resonance reminiscent of tadpole and horseshoe
orbits in the conventional restricted three-body problem. The moonlet
performs what PC called ``frog'' librations within the gap.

\begin{figure}
\begin{center}
\includegraphics[scale=1]{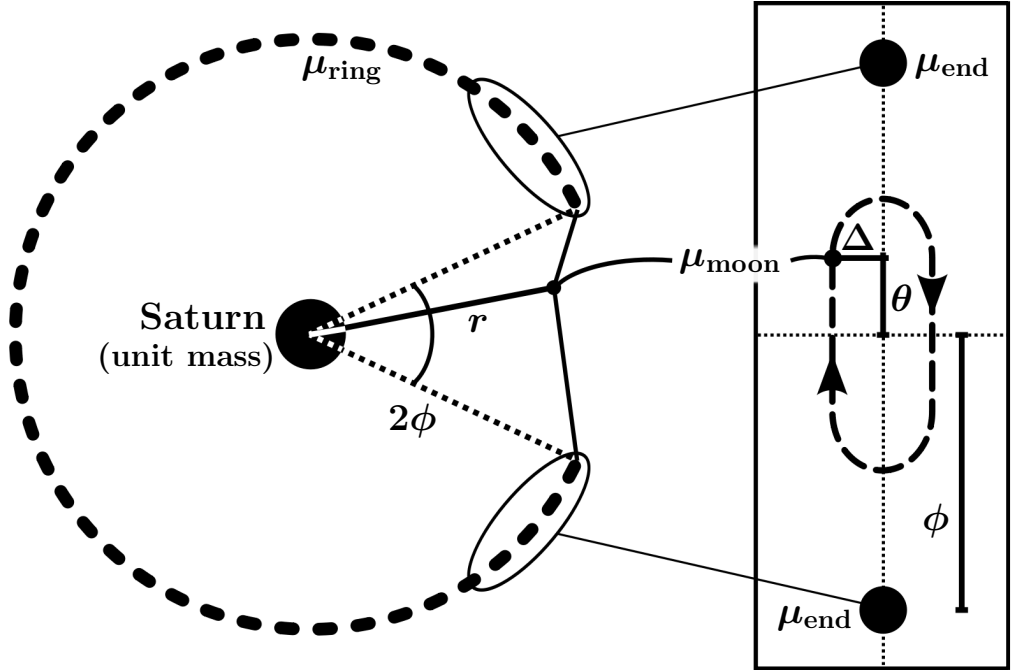}
\caption{Schematic of frog librations. The left side shows
  the geometric configuration of the moonlet and its co-orbital
  material. We use units where Saturn's mass is unity and where the
  orbit radius of the co-orbital ring material (dashed circle arc) is
  unity. The moonlet of mass $\mumoon$ located at polar coordinates
  $(r=1+\Delta,\theta)$ moves in its gap of angular size $2\phi$ in
  the co-orbital material. The portions of the co-orbital mass $\muring$
  that interact most strongly with the moonlet are those within
  azimuth $\sim$$\phi$ of the ends of the moonlet's gap. For
  simplicity, we model the entire co-orbital mass as two identical
  point masses of mass $\muend =\muring\phi/2$ located at the gap
  ends. The dashed oval curve on the right
  represents a frog libration trajectory.
\label{fig:libschematic}}
\end{center}
\end{figure}

To re-cap the main ideas behind PC's analysis we give the following
order-of-magnitude description of frog librations. Because our goal
here is to describe the physical processes clearly, we neglect factors
of order unity; we refer readers to \S\ 2 of PC for an exact
treatment. We use units where Saturn has mass $M_\mathrm{Saturn}=1$,
the co-orbital mass's distance from Saturn is $r_\mathrm{coorb}=1$,
and Newton's constant $G$ is 1. We work in the frame co-rotating with
a test particle on a circular orbit of radius 1 and (as a consequence
of our unit system) angular velocity 1. The moonlet of mass $\mumoon$
opens a gap of angular size $2\phi \ll\pi$ in the co-orbital material
(Figure~\ref{fig:libschematic}). The co-orbital material outside the
gap has mass $\muring$. Since the co-orbital material closest to the
ends of the gap interacts most strongly with the moonlet, we model the
co-orbital material as two point masses, each of mass
$\muend=\muring\phi/2$, fixed at longitudes $\pm\phi$. We label the
moonlet's polar coordinates $(r=1+\Delta,\theta)$ where $\Delta \ll 1$
and $|\theta|\ll\phi$.  The moonlet librates about the equilibrium
point%
\footnote{This equilibrium point is a local maximum
  of the gravitational potential. The moonlet can librate about this
  potential maximum because the Coriolis
  force stabilizes its motion.}  between the co-orbital masses, as shown
in the right-hand side of Figure~\ref{fig:libschematic}.

The moonlet's motion is governed by Keplerian shear
and gravitational interactions with the co-orbital masses. We
derive scaling relations for the frog libration's period and
radial/azimuthal aspect ratio as follows. The azimuthal length
$\theta_\mathrm{max}$ of one libration is of order the moonlet's azimuthal
Keplerian drift in one libration period $\Plib$.  If the libration radial
width is $\Delta_\mathrm{max}$, then the moonlet's drift speed is of the same
order, and
\begin{equation}
\theta_\mathrm{max} \sim \Plib\cdot\Delta_\mathrm{max}
\;\;\; .
\label{eqn:drift}
\end{equation}
Over the libration cycle, gravitational interactions with the
co-orbital masses change the moonlet's semimajor axis from
$1+\Delta_\mathrm{max}$ to $1-\Delta_\mathrm{max}$. The corresponding
fractional change in the moonlet's specific angular momentum is
$\sim$$\Delta_\mathrm{max}$. The angular momentum changes because the
(azimuthal) gravitational forces $\sim$$\pm\muend/\phi^2$ exerted by
the co-orbital masses do not cancel exactly; the residual is of order
$\theta/\phi$:
\begin{equation}
\Delta_\mathrm{max}
 \sim \frac{\muend}{\phi^2}\frac{\theta_\mathrm{max}}{\phi}\cdot \Plib \;\;\; .
\label{eqn:kick}
\end{equation}
Together, Eqs.~\ref{eqn:drift} and \ref{eqn:kick} imply
\begin{equation}
\Plib \sim \frac{\phi^{3/2}}{\muend^{1/2}}
\;\;\; , \;\;\;
\frac{\Delta_\mathrm{max}}{\theta_\mathrm{max}}
\sim \frac{\muend^{1/2}}{\phi^{3/2}} \;\;\; .
\label{eqn:plibsim}
\end{equation}
These scalings match those of the exact relations in Eqs.~10 and 11 of
PC, which we repeat here for reference:
\begin{equation}
\Plib = \frac{\pi}{\sqrt{3}}\frac{\phi^{3/2}}{\muend^{1/2}} 
\;\;\; , \;\;\;
\frac{\Delta_\mathrm{max}}{\theta_\mathrm{max}}
= \frac{4}{\sqrt{3}} \frac{\muend^{1/2}}{\phi^{3/2}} \;\;\; .
\label{eqn:plib}
\end{equation}
Note that the coorbital mass $\mu$ used by PC equals $2\muend$
as we define it here.

In a more detailed calculation with the co-orbital mass spread
uniformly over the entire co-orbital region outside the gap, PC
confirmed the scalings of Eq.~\ref{eqn:plibsim} and found that for
parameters characteristic of Bl\'{e}riot's environment, the
frog librations should indeed have a period of $\sim$4~years.

While the agreement between the period of the frog resonance as
estimated by PC, and the period of Bl\'{e}riot's longitude variations
as observed by \citet{tiscareno10} is promising, PC's treatment is
incomplete.  In particular, PC treated the co-orbital masses as
stationary (in the co-rotating frame).  This is a severe
simplification.  Because the moonlet creates and maintains its gap,
the co-orbital material must, to some degree, follow the moonlet's
position. If it follows too closely, the frog libration amplitude
($\theta$) may be unobservably small. On the other hand, if the gap
ends are far enough away from the moonlet, the mass there will respond
sluggishly to the moonlet's motion because of the finite time needed
for particles to drift from conjunction with the moonlet to the gap
ends. In this case, large-amplitude frog librations should be
permitted.

A self-consistent understanding of the moonlet's motion must account
for how the non-Keplerian motions of the moonlet feed back into the
non-Keplerian motions of the co-orbital masses. Here we investigate
the behavior of frog orbits when the co-orbital masses move in
response to the moonlet, and when Lindblad torques due to close
encounters between the moonlet and ring particles are present.  Our
focus here is on conceptual clarity; from a technical standpoint, our
models are, by design, crude. Many of our calculations are accurate to
order-of-magnitude at best.  In our equations we use ``$\sim$'' to
denote a relation in which order-unity factors are ignored;
``$\simeq$'' to denote a relation in which order-unity factors are
retained but some quantities remain approximate or not precisely
defined; and ``='' to denote an exact relation. In
Section~\ref{sec:feedback} we present a simple toy model for feedback
and examine its implications for frog librations.  In
Section~\ref{sec:damping} we add Lindblad torques. In
Section~\ref{sec:discussion} we summarize and comment on the results
of our experiments.

\section{Modeling the gap motion and Lindblad torques}
\label{sec:exp}

\begin{figure}
\begin{center}
\includegraphics{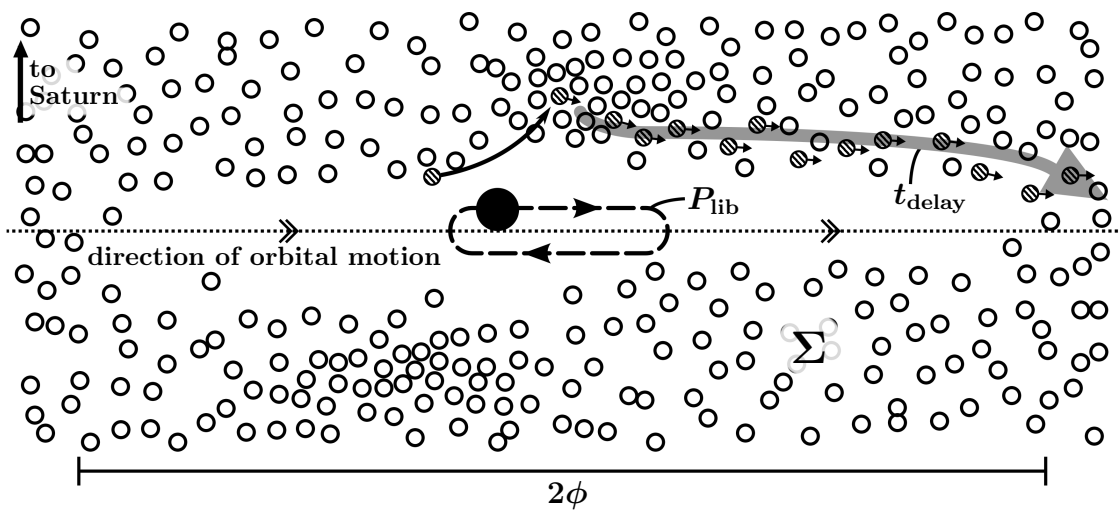}
\caption{Schematic showing the trajectory of a ring particle (shaded
  circle with arrow) participating in gap formation. A moonlet
  of mass $\mumoon$ is embedded in a particle disk of surface density
  $\Sigma$. For a particle with initial radius inside the moonlet's
  orbit (toward the top of the figure), a gravitational interaction
  with the moonlet decreases the particle's semimajor axis. Diffusion
  via collisions with nearby ring particles moves the particle back to
  its original semimajor axis over a time $\tdiff$. During this time
  the particle drifts an angle $\phi$ relative to the moonlet because
  of Keplerian shear. The dashed oval represents frog librations of
  the moonlet with period $\Plib$. \label{fig:gap}}
\end{center}
\end{figure}

Figure~\ref{fig:gap} gives an overview of the formation and
maintenance of the moonlet gap. The moonlet sits in a sea of much
smaller ring particles with mass surface density $\Sigma$. In close
encounters with ring particles, the moonlet gravitationally repels the
particles, clearing a region whose initial radial size is of order the
moonlet's Hill sphere $\mumoon^{1/3}$.  As repelled particles drift
away in longitude from the moonlet because of Keplerian shear, they
collide with other particles; over a time $\tdiff$, they diffuse back
into the gap, refilling it. Over this diffusion time, the perturbed
ring particles drift an angular distance $\phi$ from their close
encounter with the moonlet to the end of the gap. That is,
\begin{equation}
\tdiff \simeq \frac{2}{3} \frac{\phi}{x_\mathrm{gap}}
\label{eqn:tdiff}
\end{equation}
where $x_\mathrm{gap} < \mumoon^{1/3}$ is the gap's typical downstream
radial half-width (radial distance between the moonlet's orbit and one
edge of the gap). In this definition $x_\mathrm{gap}$ is the gap width
along most of the gap's (azimuthal) length, far downstream of the moonlet's
location; its value depends on the way ring particles diffuse back into the
gap and, therefore, on the local ring viscosity. The downstream gap width
$x_\mathrm{gap}$ may differ from the gap width
$x\sim\mumoon^{1/3}$ at the moonlet's longitude.

We can think of $\tdiff$ as the delay between any movement of the
moonlet and the response of the gap ends to that movement. 
If the moonlet were to suddenly shift its position in the
rotating frame, the gap ends would re-position themselves so as to be
centered on the moonlet's new position---but only after a time
$\tdiff$. This simple picture suggests that we modify the model of
PC so that at any given instant, the co-orbital masses are
located $\pm\phi$ in longitude away from the moonlet's location a time
$\tdiff$ ago.

\subsection{Experiments including gap motion only}
\label{sec:feedback}

To explore the effects of this delayed gap motion on the moonlet, we
construct a numerical toy model. As in PC, we treat the co-orbital
material as two point masses each of mass $\muend$ located at the
ends of the moonlet gap, and we integrate the trajectory of the
moonlet (test particle) as it interacts with the point masses and the
central unit mass. To model the feedback, we fix the point masses'
angular separation to be $2\phi$ and their radial coordinate to be
unity, and force them to move so that the longitude of their midpoint
at time $t$ equals the moonlet's longitude at time $t-\tdiff$. To
begin the calculation, we integrate from $t=0$ to $t = \tdiff$ with
stationary point masses, and then use this history when we turn on the feedback
at time $t = \tdiff$. We treat $\muend$, $\phi$, and $\tdiff$ as free
parameters.

\begin{figure}
\begin{center}
\includegraphics[scale=.5]{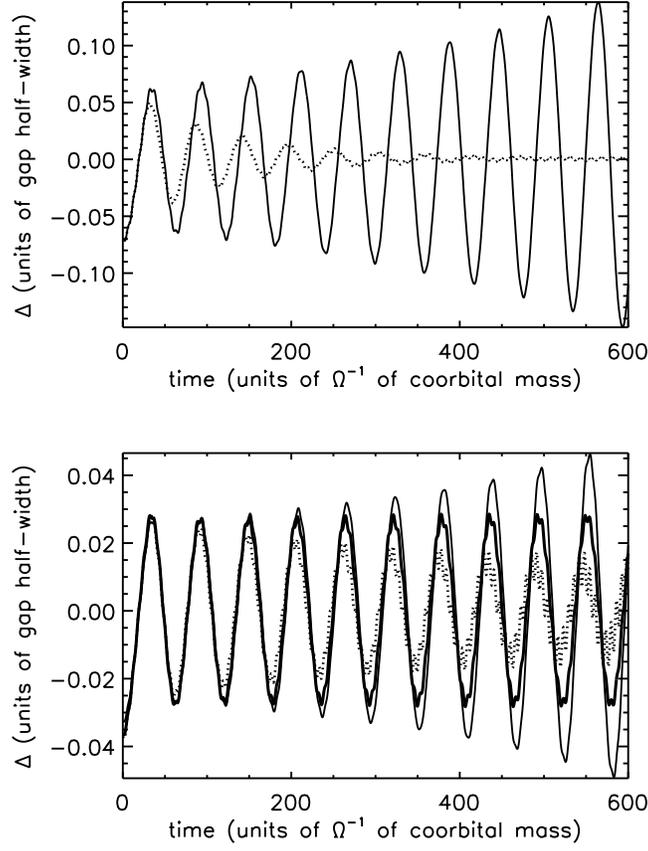}
\caption{Sample moonlet trajectories showing the effects of feedback
  from co-orbital mass motion and damping from Lindblad torques. The
  parameter values $\muend = 10^{-4.5}$ and $\phi=0.4$ are used
  throughout, so that the nominal libration period according to
  Eq.~\ref{eqn:plib} is $P_\mathrm{lib} = 81.6$. Including
  co-orbital mass motion changes the potential the moonlet sees; in
  all the sample trajectories shown here, the libration period in the
  altered potential is actually $\Plib\simeq 60$. The top panel shows
  the effects of co-orbital mass motion alone with different values of
  $\tdiff$, the time by which the co-orbital mass motion is delayed
  relative to the moonlet's motion (i.e., the gap re-adjustment time).
  Values of $\tdiff$ greater than some threshold lead to increasing
  libration amplitudes (solid curve, $\tdiff = 30$).  Values of
  $\tdiff$ even slightly smaller than the threshold produce decreasing
  amplitudes (dotted curve, $\tdiff=25$). The bottom panel includes
  the effects of both gap movement and Lindblad disk torques. All
  three trajectories have $\tdiff=30$ and $\Sigma=0.01$, but the
  moonlet masses are $\mumoon=\{1.48,7.41,37.1\}\times 10^{-7}$,
  respectively, for the light solid, heavy solid, and dotted curves,
  and the gap widths are $x = 6$ moonlet Hill radii. The coefficient
  of 6 is given approximately by the edges of the propeller gaps
  computed by \cite{spahn00}. The trajectories shown here are
  representative of our numerical experiments and are meant to
  illustrate our findings; the parameter values used do not correspond
  to the propeller Bl\'{e}riot. \label{fig:feedback}}
\end{center}
\end{figure}

The top panel of Figure~\ref{fig:feedback} shows representative
results from this toy model. When $\tdiff$ is short compared to the
nominal frog libration time (Eq.~\ref{eqn:plib}), the librations damp
quickly. When $\tdiff$ is comparable to $\Plib$ or longer, the
libration amplitude grows and the moonlet eventually exits the
resonance. The transition between amplitude growth and amplitude decay
occurs when $\Plib$ is a few times $\tdiff$; in the representative
example shown in Figure~\ref{fig:feedback}, this transition occurs
when $\tdiff\sim\Plib/3$.

The shape of the potential the moonlet experiences also
changes because of the co-orbital masses' motion. The change in the
potential alters $\Plib$: in
Figure~\ref{fig:feedback}, $\Plib\simeq 60$ rather than 
$\Plib\simeq 81.6$ as predicted by Eq.~\ref{eqn:plib}. In our numerical trials
using parameter values spanning the ranges $10^{-7}<\muend<10^{-4}$,
$10^{-3}<\phi<0.5$, and $10<\tdiff<7000$, we found that the ratio between
$\tdiff$ and $\Plib$ is the principal deciding factor for the
moonlet's qualitative behavior.

We interpret the growing librations when $\tdiff \gtrsim \Plib$ as
resonant forcing. In this case, the gap position lags the moonlet
position by a phase of order unity or, equivalently, a longitude of
order the libration amplitude. Because the co-orbital masses are
slaved to the moonlet's actual past motion, they automatically drive the
moonlet at resonance.\footnote{Since the co-orbital masses follow the
  moonlet's past rather than its current motion, the driving frequency
  may, in principle, differ slightly from resonance, particularly for
  $\tdiff\gg\Plib$. However, the driving frequency always moves toward
  resonance with time.}

In the limit $\tdiff\ll\Plib$, however, the moonlet's displacement
from the gap center is always small compared to the libration
amplitude that a moonlet with the same location and velocity would
have if the co-orbital masses were stationary. The moonlet's speed
relative to the gap is also much less than the speed it would have if
the co-orbital masses were stationary. Continuously shrinking the
moonlet's displacement from the equilibrium point between the
co-orbital masses, and its speed relative to the equilibrium point,
damps the libration amplitude until the moonlet lands at the fixed
point of the potential.

We find that the transition between the regimes of growing and
decaying libration amplitude occurs over a narrow range in $\tdiff$. For
example, the values for $\tdiff$ corresponding to the two different
outcomes in the top panel of Figure~\ref{fig:feedback} differ by less
than 10\% of $\Plib$.

\begin{figure}
\begin{center}
\includegraphics[scale=.6]{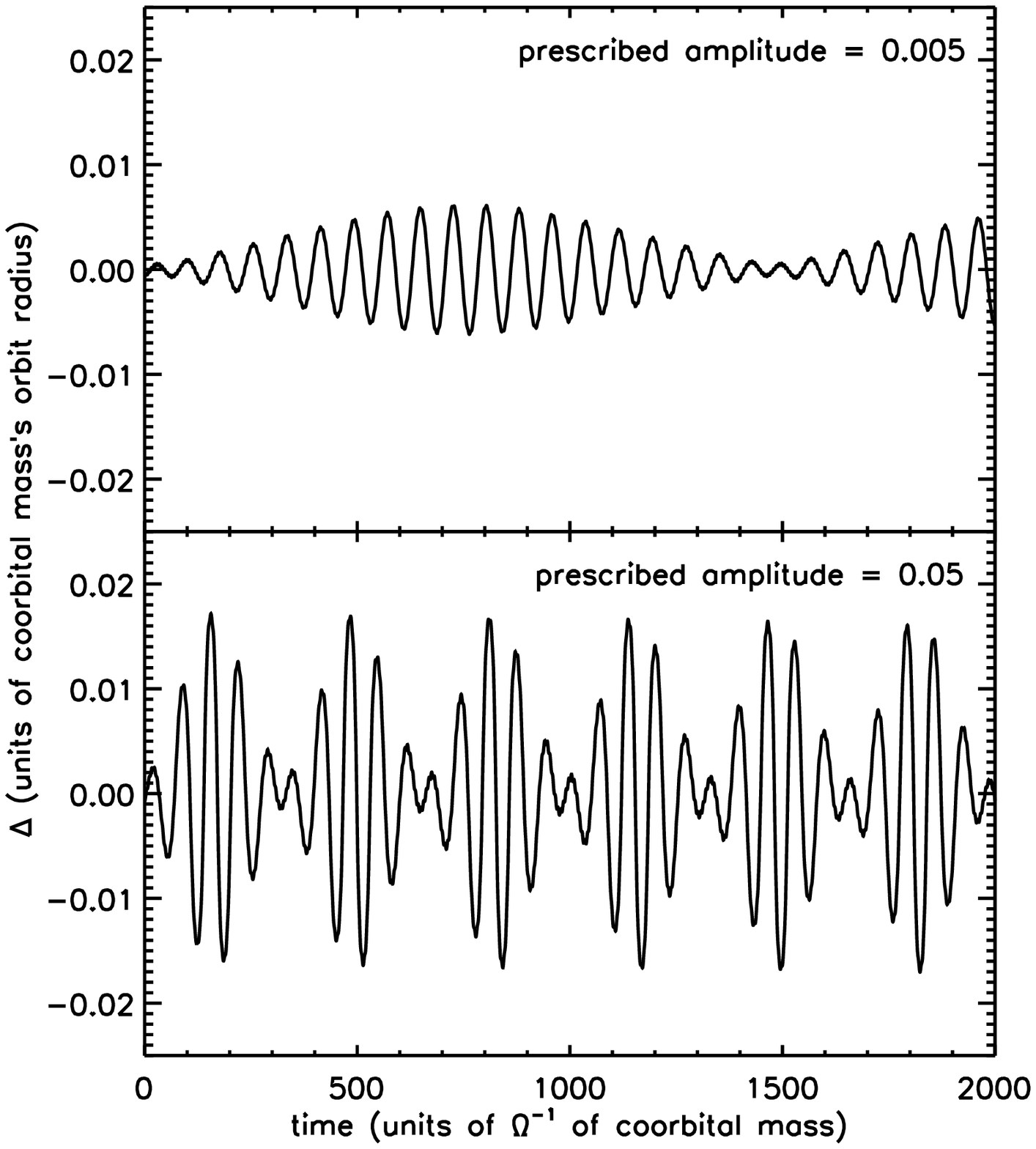}
\caption{Sample moonlet trajectories from experiments where the
  co-orbital masses move sinusoidally with prescribed amplitude and
  period. For both panels, the prescribed period is 81.6 (equal
  to the period that the moonlet would have if the co-orbital masses
  were stationary). The top panel shows the
  moonlet's motion when the co-orbital masses move with amplitude 0.005
  and the bottom panel shows results when that amplitude is 0.05. The
  stronger driving in the bottom panel trajectory leads to both a
  larger beat amplitude and a higher beat frequency. The latter
  implies that the co-orbital masses' motion changes the potential seen
  by the moonlet and the natural frequency of oscillations in
  that potential. \label{fig:prescr}}
\end{center}
\end{figure}

We performed some supplementary experiments. First we forced the
co-orbital masses to move sinusoidally in longitude with prescribed
amplitude and frequency. Setting the co-orbital masses' oscillation
period equal to $\Plib$ as given by Eq.~\ref{eqn:plib} produces beats
in the moonlet's motion: the motion of the co-orbital masses changes
the shape of the potential the moonlet sees, shifting the libration
period enough for the driving to be just off resonance
(Figure~\ref{fig:prescr}). If we decrease the co-orbital masses'
oscillation amplitude while keeping their oscillation period fixed,
the potential experienced by the moonlet changes less.  Then the
driving is closer to resonance and the beat frequency
decreases. Compare these results with those of our original
experiments, where the co-orbital masses were slaved to the moonlet's
actual past motion and the driving was therefore exactly on resonance.

In experiments aimed at softening the transition between the regimes
of growing and decaying amplitude, we performed simulations where we
artificially weakened the feedback strength. To do this we forced the
co-orbital masses to follow the moonlet's past motion, as before, but
scaled the amplitude of the co-orbital masses' motions relative to the
moonlet's by a constant factor. We observed no qualitative change in
the moonlet's behavior. The value of $\tdiff$ at the transition
changed by a factor of order unity.

To summarize our findings so far: the libration amplitude damps to
zero if $\tdiff$ is too short and grows without bound if $\tdiff$ is
comparable to or longer than $\Plib$, with an abrupt transition
between the two regimes.  However, longitude residuals like those
observed might be produced in the latter regime if some other
mechanism damps the motion enough to limit the libration amplitude's
growth. We explore one possible damping mechanism in
\S\ref{sec:damping}.

The requirement that $\tdiff\gtrsim\Plib/3$ that we find in
our simulations (the factor of 3 is empirical)
has a simple consequence for the gap's radial width:
using Eq.~\ref{eqn:plib} and $\muend \simeq \Sigma\cdot
2x_\mathrm{gap}\cdot \phi$, we find that
\begin{equation}
\tdiff \gtrsim \frac{\pi}{3\sqrt{3}}\frac{\phi^{3/2}}{\muend^{1/2}}
       \simeq \frac{\pi}{3\sqrt{6}}\frac{\phi}{\sqrt{\Sigma x_\mathrm{gap}}}
\label{eqn:tdiffsigma} 
\end{equation}
which together with Eq.~\ref{eqn:tdiff} implies that
\begin{equation}
\frac{x_\mathrm{gap}}{r_\mathrm{coorb}}
\lesssim 
\frac{24}{\pi^2}\frac{\Sigma r_\mathrm{coorb}^2}{M_\mathrm{Saturn}}
\label{eqn:lib_eq_diff}
\end{equation}
or
\begin{equation}
x_\mathrm{gap} 
       \lesssim 4\;\m\, \left( \frac{\Sigma}{40\;\g/{\rm cm}^2}  \right)
                \left(\frac{r_\mathrm{coorb}}{1.34\times 10^{10}\;\cm}\right)^3
\label{eqn:lib_eq_diff_num}
\end{equation}
where for the last line we have restored the units and used numbers inspired
by Bl\'{e}riot and the A ring \citep{tiscareno10,colwell09}.  

At face value, Eq.~\ref{eqn:lib_eq_diff_num} suggests that for
Bl\'{e}riot's moonlet to avoid libration damping, its gap should be
radially narrow, of order several meters across, over a significant
portion of its azimuthal extent. The required narrowness seems
implausible, as it is comparable in size to the meter-sized particles dominating the mass in the A ring
\citep[see Table 15.1 of][]{cuzzi09}.  Thus,
Eq.~\ref{eqn:lib_eq_diff_num} argues against frog librations as being
the correct explanation for the observed non-Keplerian motions of
propellers.  We stress, however, that our equations are only as
accurate as the toy model for $\tdiff$ from which they derive. A more 
careful analysis of the gap's reactions to the moonlet 
(e.g., using numerical simulations of the entire gap) would provide
an important test of our estimates here.

\subsection{Experiments including gap motion and Lindblad torques}
\label{sec:damping}

The results of \S\ref{sec:feedback} indicate that stable frog
librations can persist only if $\tdiff\gtrsim \Plib$ and if some other
mechanism damps the librations enough to stop their runaway growth.
One damping mechanism is disk torques due to
Lindblad resonances---the same torques that open the gap and are
responsible for Type I migration.  Because Lindblad torques repel the
moonlet away from the gap edges, forcing the moonlet's semimajor axis
back toward the gap center, we expect such torques to decrease the
frog libration amplitude.

To provide a conceptual basis for the behavior we expect in our
numerical experiments, we give the following order-of-magnitude
estimates of the Lindblad torques and associated damping timescale,
dropping factors of order unity.  According to the standard
impulse approximation (see, for example, \citet{dermott84} for a
derivation using the impulse approximation or
\citet{goldreichtremaine82} for a derivation using Lindblad
resonances), the moonlet's azimuthal acceleration due to its
``one-sided'' interaction with ring material on one side of its gap is
of order
\begin{equation}
F_\mathrm{\theta, 1-sided} 
\sim \frac{\mumoon\Sigma}{x^3} \;\;\; .
\end{equation}
This is consistent with the scaling $\Delta J\propto \mumoon^2/x^5$
for the one-sided impulse of angular momentum per unit disk mass
received by the moonlet (see, e.g., Eq.~32 of \cite{cridaetal}:
multiply their equation by the disk mass $\sim$$\Sigma x^2$ that
passes conjunction with the moonlet per unit time --- this is
analogous to the integral over $x$ that they describe with their
Eq.~36 --- and divide by $\mumoon$ to get an acceleration rather than
a force).  Note that we use $x$ here rather than $x_{\rm gap}$ because
we are concerned with ring material that just passes conjunction with
the moonlet, i.e., ring material near the moonlet's longitude.

The moonlet interacts
simultaneously with ring material radially interior and exterior to
it.  If we assume the ring surface density is uniform, then the two
one-sided torques cancel up to any asymmetry in the moonlet's radial
distances to the gap edges. Then the moonlet's net acceleration is
\begin{equation}
F_\mathrm{\theta} \sim -\frac{\Delta}{x} \cdot F_\mathrm{\theta, 1-sided}
\sim -\frac{\mumoon\Sigma\Delta}{x^4} \;\;\; .
\label{eqn:ftheta}
\end{equation}
The minus sign enters because the moonlet is repelled more by
whichever gap edge is closer to it. 
The damping timescale is 
\begin{equation}
t_\mathrm{damp} \sim \left| \frac{\dot\theta}{F_\mathrm{\theta}} \right| \sim \frac{x^4}{\mumoon\Sigma} \;\;\; 
\label{eqn:tdamp}
\end{equation}
where we have used $|\Delta| \sim |\dot\theta|$ due to Keplerian shear 
(see Eq.~\ref{eqn:drift}).

We combine Lindblad damping with co-orbital feedback by adding the
term $F_\mathrm{\theta}$ (as given by Eq.~\ref{eqn:ftheta}, with an
assumed coefficient of unity) to the moonlet's azimuthal equation of
motion in the numerical toy model of Section~\ref{sec:feedback}. We
expect that damping will balance the (positive) feedback when $t_{\rm
  damp}$ is of order the libration growth timescale. In the example
shown in the top panel of Figure~\ref{fig:feedback}, the timescale to
amplify librations is about $t_\mathrm{amplify} = 6$ libration periods
of the undriven system (driving alters the potential and shortens the
libration period). To check quickly that we understand our simulations
in the bottom panel of that figure, we make a simple
order-of-magnitude estimate.  We set $t_\mathrm{damp} =
t_\mathrm{amplify} = 6\Plib$ and substitute Eqs.~\ref{eqn:plib},
\ref{eqn:tdamp}, and the parameters of the example in
Figure~\ref{fig:feedback}: $\muend=10^{-4.5}$, $x=6(\mumoon/3)^{1/3}$
or six Hill radii of the moonlet, $\Sigma = 0.01$, and $\phi=0.4$. In
this example, damping balances growth when $\mumoon\sim
(3^{5/2}\pi^3/6^9)\cdot \phi^{9/2}\Sigma^3\muend^{-3/2}\sim 4\times
10^{-6}$. Our simulations show that a somewhat smaller $\mumoon\sim
7.5\times 10^{-7}$ is needed (bottom panel of
Figure~\ref{fig:feedback}), but the agreement is reasonable given the
order-of-magnitude nature of our arguments.

Since our Eqs.~\ref{eqn:ftheta} through \ref{eqn:tdamp} neglect all
order-unity coefficients --- strictly speaking, they are scaling
relations --- we may apply them to real systems only with the
understanding that numerical results cannot be more than broadly
suggestive.  If we use $x\sim 6$ moonlet Hill radii as before, then
Eqs.~\ref{eqn:plib}, \ref{eqn:tdamp}, and the equality in
Eq.~\ref{eqn:lib_eq_diff} ($x_\mathrm{gap}/\Sigma\sim 24/\pi^2$) give
\begin{equation}
\frac{t_\mathrm{damp}}{\Plib} 
\sim 0.6
     \left(\frac{\mumoon}{3\times 10^{-15}}\right)^{1/3}
     \left(\frac{\phi}{0.004}\right)^{-1}
\end{equation}
where the numerical values are inspired by Bl\'{e}riot
\citep{tiscareno10}.  This numerical result suggests that positive
co-orbital feedback and damping Lindblad torques might plausibly
balance for propellers in the outer A ring.  Unfortunately, the result
relies on Eq.~\ref{eqn:lib_eq_diff}, which as we have seen previously
demands an unrealistically narrow gap.

\section{Summary and Discussion}
\label{sec:discussion}

As a step toward a self-consistent model for propellers' non-Keplerian
motions, we added to the frog libration model of PC new terms
representing co-orbital mass motion and Lindblad torques.  In
numerical experiments with this `extended' frog model, we found that
allowing the co-orbital mass at the ends of the gap to move in
response to the moonlet's non-Keplerian motions can either damp the
frog librations completely or drive the librations resonantly.  In
other words, in our simple model for co-orbital feedback, the feedback
is either strongly negative or strongly positive.  We found further
that the strong positive feedback could be limited, and the motion
stabilized, by Lindblad torques.

Our numerical experiments emphasize simplicity over realism.  Although
they clarify the conditions necessary for stable frog librations, they
offer no explanation for why these conditions should be met. That is,
we leave unanswered the question of why $\tdiff$, $\Plib$,
$\mumoon$, and $\muring$ of real propeller systems should occur in the
right proportions for stable frog librations to exist.

Still, our extended frog model makes observationally testable
predictions: if propeller moonlets are performing frog librations, we
expect that 
1) the properties
of Bl\'eriot's gap (its azimuthal and radial dimensions, and the co-orbital
surface density outside the gap) are such that its frog libration period
$\Plib \simeq 4$ yr;
2) Bl\'{e}riot's and other propellers' longitude
residuals will continue to vary sinusoidally in time;
3) at any
given time, propeller positions will be typically offset from the
centers of their gaps by azimuthal distances of order the observed rms
longitude residuals; and
4) the properties of propeller gaps are such that
gap drift (equivalently, gap closing) timescales $\tdiff$
(Eq.~\ref{eqn:tdiff}) are comparable to or longer than frog
libration periods $\Plib$.
Of these four predictions, the first three are retained from
PC's original frog model, while the fourth is new from
the extended frog model.

Unfortunately, we cannot seem to satisfy both predictions 1 and 4 for
Bl\'{e}riot.  If we take the azimuthal length of Bl\'eriot's gap to be
$\phi \approx 0.004$ and its radial width $x_{\rm gap}$ to be a few km
(a few moonlet Hill radii), we can indeed reproduce $\Plib \simeq 4$
yr, in accord with the {\it Cassini} observations (PC, see the
discussion following their Eq.~12).  But a gap of such dimensions
would close in a time $\tdiff \sim 0.01 \Plib$ (Eq.~\ref{eqn:tdiff})
and would lead to a libration amplitude much smaller than the 260-km
amplitude that is observed.  For the gap drift time $\tdiff$ to be
longer than the frog libration time $\Plib$, the radial width of the
gap may need to be unrealistically small --- comparable to the ring
particle size. This conclusion is tentative because our treatment of
feedback is primitive and possibly oversimplified.  We state this
shortcoming of our model in the hope that more accurate
studies---e.g., numerical simulations designed to explore long-range
interactions between the moonlet and the entire gap---may either
confirm that the frog resonance is not responsible for the observed
non-Keplerian motions, or reveal that feedback works in a way
different from what we have imagined in this paper.

\acknowledgments We thank an anonymous referee for a detailed reading
of this work.


\begin{thebibliography}{14}
\expandafter\ifx\csname natexlab\endcsname\relax\def\natexlab#1{#1}\fi

\bibitem[{Colwell {et~al.}(2009)Colwell, Nicholson, Tiscareno, Murray, G., \&
  Marouf}]{colwell09}
Colwell, J.~E., Nicholson, P.~D., Tiscareno, M.~S., Murray, C.~D., G., F.~R.,
  \& Marouf, E.~A. 2009, in {\em Saturn from Cassini-Huygens}, ed.
  M.~Dougherty, L.~Esposito, \& T.~Krimigis (Heidelberg: Springer), 375--412

\bibitem[{{Crida} {et~al.}(2010){Crida}, {Papaloizou}, {Rein}, {Charnoz}, \&
  {Salmon}}]{cridaetal}
{Crida}, A., {Papaloizou}, J.~C.~B., {Rein}, H., {Charnoz}, S., \& {Salmon}, J.
  2010, \aj, 140, 944

\bibitem[{{Cuzzi} {et~al.}(2009){Cuzzi}, {Clark}, {Filacchione}, {French},
  {Johnson}, {Marouf}, \& {Spilker}}]{cuzzi09}
{Cuzzi}, J., {Clark}, R., {Filacchione}, G., {French}, R., {Johnson}, R.,
  {Marouf}, E., \& {Spilker}, L. 2009, {Ring Particle Composition and Size
  Distribution}, ed. {Dougherty, M.~K., Esposito, L.~W., \& Krimigis, S.~M.},
  459--+

\bibitem[{{Dermott}(1984)}]{dermott84}
{Dermott}, S.~F. 1984, {Dynamics of narrow rings}, ed. {R.~Greenberg \&
  A.~Brahic}, 589--637

\bibitem[{{Goldreich} \& {Tremaine}(1982)}]{goldreichtremaine82}
{Goldreich}, P., \& {Tremaine}, S. 1982, \araa, 20, 249

\bibitem[{{Lewis} \& {Stewart}(2009)}]{lewis09}
{Lewis}, M.~C., \& {Stewart}, G.~R. 2009, \icarus, 199, 387

\bibitem[{{Pan} \& {Chiang}(2010)}]{pan10}
{Pan}, M., \& {Chiang}, E. 2010, \apjl, 722, L178

\bibitem[{{Sei{\ss}} {et~al.}(2005){Sei{\ss}}, {Spahn}, {Srem{\v c}evi{\'c}},
  \& {Salo}}]{sei05}
{Sei{\ss}}, M., {Spahn}, F., {Srem{\v c}evi{\'c}}, M., \& {Salo}, H. 2005,
  \grl, 32, 11205

\bibitem[{{Spahn} \& {Srem{\v c}evi{\'c}}(2000)}]{spahn00}
{Spahn}, F., \& {Srem{\v c}evi{\'c}}, M. 2000, \aap, 358, 368

\bibitem[{{Srem{\v c}evi{\'c}} {et~al.}(2007){Srem{\v c}evi{\'c}}, {Schmidt},
  {Salo}, {Sei{\ss}}, {Spahn}, \& {Albers}}]{srem07}
{Srem{\v c}evi{\'c}}, M., {Schmidt}, J., {Salo}, H., {Sei{\ss}}, M., {Spahn},
  F., \& {Albers}, N. 2007, \nat, 449, 1019

\bibitem[{{Srem{\v c}evi{\'c}} {et~al.}(2002){Srem{\v c}evi{\'c}}, {Spahn}, \&
  {Duschl}}]{srem02}
{Srem{\v c}evi{\'c}}, M., {Spahn}, F., \& {Duschl}, W.~J. 2002, \mnras, 337,
  1139

\bibitem[{{Tiscareno} {et~al.}(2008){Tiscareno}, {Burns}, {Hedman}, \&
  {Porco}}]{tiscarenoetal08}
{Tiscareno}, M.~S., {Burns}, J.~A., {Hedman}, M.~M., \& {Porco}, C.~C. 2008,
  \aj, 135, 1083

\bibitem[{{Tiscareno} {et~al.}(2006){Tiscareno}, {Burns}, {Hedman}, {Porco},
  {Weiss}, {Dones}, {Richardson}, \& {Murray}}]{tiscarenoetal06}
{Tiscareno}, M.~S., {Burns}, J.~A., {Hedman}, M.~M., {Porco}, C.~C., {Weiss},
  J.~W., {Dones}, L., {Richardson}, D.~C., \& {Murray}, C.~D. 2006, \nat, 440,
  648

\bibitem[{{Tiscareno} {et~al.}(2010){Tiscareno}, {Burns}, {Srem{\v c}evi{\'c}},
  {Beurle}, {Hedman}, {Cooper}, {Milano}, {Evans}, {Porco}, {Spitale}, \&
  {Weiss}}]{tiscareno10}
{Tiscareno}, M.~S., {Burns}, J.~A., {Srem{\v c}evi{\'c}}, M., {Beurle}, K.,
  {Hedman}, M.~M., {Cooper}, N.~J., {Milano}, A.~J., {Evans}, M.~W., {Porco},
  C.~C., {Spitale}, J.~N., \& {Weiss}, J.~W. 2010, \apjl, 718, L92

\end{thebibliography}

\end{document}